\documentclass[letterpaper, 10 pt,conference]{ieeeconf}  
\usepackage{amssymb}
\usepackage{amsmath,mathrsfs}
\usepackage{color}
\usepackage{graphicx}
\usepackage{mathtools}
\usepackage{lineno,hyperref}
\usepackage{cleveref}
\usepackage{bm}
\let\labelindent\relax
\usepackage{enumitem}
\usepackage{comment}
\usepackage{tikz}
\usepackage{setspace}
\usepackage{algorithm}
\usepackage{algorithmicx}
\usepackage{algpseudocode}
\modulolinenumbers[5]
\newtheorem{theorem}{Theorem}[section]
\newtheorem{remark}{Remark}[section]

\newtheorem{corollary}{Corollary}[section]

\newtheorem{assumption}{Assumption}[section]

\usepackage{multirow}
\usepackage{color}
\usepackage{cite}
\newcommand{\hop}{\mathsf{H}}

\usepackage{eso-pic}
\AddToShipoutPictureBG*{%
	\AtPageUpperLeft{%
		\setlength\unitlength{1in}%
		\hspace*{\dimexpr0.5\paperwidth\relax}
		\makebox(-0.18,-1)[c]{
			\begin{tabular}{c c}
				Rodrigo A. Gonz\'alez \emph{et al.},
				Sampling in Parametric and Nonparametric System Identification: Aliasing, Input Conditions, and Consistency. \\
				To appear in
				{\em IEEE Control Systems Letters (L-CSS)},
				2024,
				uploaded to arXiv on October 25th, 2024. \\
\end{tabular}}}}

\providecommand{\keywords}[1]{\textbf{\textit{Index terms---}}}

\IEEEoverridecommandlockouts                              
\title{\LARGE \bf Sampling in Parametric and Nonparametric System Identification: Aliasing, Input Conditions, and Consistency}

\author{Rodrigo A. Gonz\'alez, Max van Haren, Tom Oomen and Cristian R. Rojas 
\thanks{This work was partly supported by the Swedish Research Council under contract number 2023-05170, and by the ECSEL Joint Undertaking under grant agreement 101007311 (IMOCO4.E). The Joint Undertaking receives support from the European Union Horizon 2020 research and innovation programme.}%
\thanks{R. A. Gonz\'alez, M. van Haren and T. Oomen are with the Control Systems Technology Section, Department of Mechanical Engineering, Eindhoven University of Technology, Eindhoven, The Netherlands. T. Oomen is also with the Delft Center for Systems and Control, Delft University of Technology, Delft, The Netherlands. C. R. Rojas is with the Division of Decision and Control Systems, KTH Royal Institute of Technology, Stockholm, Sweden. E-mail of the corresponding author: r.a.gonzalez@tue.nl.}%
}
\begin{document}

\maketitle

\begin{abstract}
The sampling rate of input and output signals is known to play a critical role in the identification and control of dynamical systems. For slow-sampled continuous-time systems that do not satisfy the Nyquist-Shannon sampling condition for perfect signal reconstructability, careful consideration is required when identifying parametric and nonparametric models. In this letter, a comprehensive statistical analysis of estimators under slow sampling is performed. Necessary and sufficient conditions are obtained for unbiased estimates of the frequency response function beyond the Nyquist frequency, and it is shown that consistency of parametric estimators can be achieved even if input frequencies overlap after aliasing. Monte Carlo simulations confirm the theoretical properties.   
\end{abstract}
\begin{keywords}
Frequency-domain system identification; Undersampled systems; Frequency response function.
\end{keywords}

\section{Introduction}
\label{sec:introduction}

The exact reconstruction of continuous-time signals based on their samples is crucial when designing data-driven methods for estimating dynamical systems. According to the Nyquist-Shannon theorem, such reconstruction is possible if and only if the sampling rate is at least twice the bandwidth of the signal \cite{oppenheim1997signals}. If the Nyquist-Shannon condition is not met, there exists an overlap of frequency components, known as aliasing. Since aliasing causes a loss of information for frequencies above the Nyquist frequency, system identification methods typically minimize the reconstruction error by using a small sampling period to ensure that the relevant dynamics remain below the Nyquist frequency \cite{ljung1998system}.

When small sampling periods are not feasible, one way to mitigate aliasing effects is by multi-rate system identification. Methods have been developed to identify fast-dynamics models in multi-rate settings by exploiting the fast sampling of inputs \cite{zhu2009system,liu2014efficient}. Polynomial transformation techniques \cite{ding2010modified} and state-space approaches \cite{yin2020subspace} for dual-rate systems have also been proposed. These methods are inherently designed for discrete-time multi-rate models, and do not directly address continuous-time systems under slow sampling conditions.

For uniformly-sampled continuous-time systems, evidence suggests that slow sampling does not necessarily pose a fundamental limitation for the identification of stochastic differential equations \cite{aastrom1969choice}. However, slow sampling can degrade performance of transfer function identification methods due to inaccuracies in estimating output derivatives \cite{ljung2009experiments}. Moreover, existing statistical analyses \cite{pan2020consistency,gonzalez2020consistent} assume that the sampling period is fast enough to allow for a bijective transformation between continuous-time and discrete-time models \cite{kollar1996equivalence}, which may not hold in slow sampling scenarios.

Several methods for frequency-response function estimation beyond the Nyquist frequency have been developed. A practical nonparametric identification method beyond the Nyquist frequency is proposed in \cite{ehrlich1989identification}, which requires careful selection of input signals to avoid aliasing. Similarly, \cite{van2023beyond} develops a local polynomial approach to allow for disentanglement of the aliased contributions for slow-sampled systems assuming a smooth frequency response. While  promising, these methods lack a rigorous statistical foundation, such as explicit conditions on the input signals for well-posedness of the methods, and optimality guarantees.

Although important developments have been made, a comprehensive statistical analysis of continuous-time system identification methods when the sampling violates the Nyquist-Shannon condition is still needed. This gap is addressed through our main contributions, which are as follows.

\begin{enumerate}[label=C\arabic*]
	\item
	\label{contribution1}
    We prove that the least-squares estimator of the frequency response function is unbiased even when the Nyquist-Shannon sampling condition is violated, provided the input frequencies remain distinct after accounting for aliasing. We compute the covariance of this estimator, and provide an alternative interpretation of it as the quotient of output and input discrete-time Fourier transforms, evaluated at the input frequencies.
    \item 
    \label{contribution2}
    Using the least-squares estimator analyzed in Contribution \ref{contribution1}, we provide an explicit relation, valid for finite sample size, between frequency domain and time domain identification in the slow-sampling regime.
    \item
    \label{contribution3}
    We prove that the time-domain prediction error method is a consistent estimator of the parametric model under mild identifiability conditions, even if the input frequency lines are aliased, or if they overlap after aliasing.
\end{enumerate}
The remainder of this letter is organized as follows. Section~\ref{sec:problemformulation} introduces the identification setup. Section \ref{sec:leastsquares} presents the unbiased least-squares frequency response estimator, and Section \ref{sec:parametric} links this estimator to parametric estimators and provides conditions for consistency. Simulation results are provided in Section \ref{sec:simulations}, and Section \ref{sec:conclusions} concludes this letter.
  
\textit{Notation}: The Heaviside operator satisfies $p\mathbf{x}(t)=\frac{d\mathbf{x}(t)}{dt}$. All vectors, matrices, and multivariate signals are written in bold, and vectors are column vectors, unless transposed. The conjugate transpose of a vector $\mathbf{x}$ is denoted as $\mathbf{x}^\hop$, and the diagonal matrix $\textnormal{diag}(\mathbf{x})$ or $\textnormal{diag}(x_1,\dots,x_n)$ is formed by the elements in $\mathbf{x}=[x_1,\dots,x_n]^\top$. The identity matrix of size $m$ is denoted as $\mathbf{I}_m$, and the null space of a matrix $\mathbf{A}$ is denoted as $\textnormal{Ker}(\mathbf{A})$. The Kronecker delta centered at $n=0$ is denoted as $\delta(n)$, and $b \textnormal{ mod }m$ denotes the modulo operation. The expression $\overline{\mathbb{E}}\{x(kh)\}=\lim_{N\to\infty}\frac{1}{N}\sum_{k=1}^N \mathbb{E}\{x(kh)\}$ denotes the expectation of a quasi-stationary signal $x(kh)$. 

\section{Problem formulation}
\label{sec:problemformulation}
Consider a single-input single-output, linear time-invariant (LTI), asymptotically stable, continuous-time system
\begin{equation}
\label{system}
    x(t) = G_0(p)u(t) ,
\end{equation}
where $x(t)\in\mathbb{R}$ is the output of the LTI system $G_0(p)$ when subject to the input $u(t)\in\mathbb{R}$. We are interested in estimating $G_0(p)$ from input and output data. Thus, we excite the system with a continuous-time multisine input of the form
\begin{equation}
\label{input}
    u(t)=a_{0} + \sum_{\ell =1}^M a_\ell \cos(\omega_\ell t + \phi_{\ell}),
\end{equation}
where the angular frequencies $\omega_\ell$ are positive and sorted in strictly increasing order, and $\{a_\ell\}_{\ell=0}^M$ are all different from zero. We assume perfect knowledge of the continuous-time input signal and the absence of a hold device between the input source and the system, which implies that $u(t)$ is a band-limited signal with bandwidth $\omega_M$.

We retrieve a noisy measurement of the output $x(t)$ at equidistant time points $t=h,2h,\dots,Nh$:
\begin{equation}
\label{output}
    y(kh)=x(kh)+v(kh), \quad k=1,\dots,N,
\end{equation}
where $v(kh)\in\mathbb{R}$ is a zero-mean i.i.d. white noise of variance $\sigma^2$. The key assumption in this work is that the sampling period $h$ does not satisfy the Nyquist-Shannon criterion for perfect reconstructability of the input signal based only on its samples, i.e., $h\geq \pi/\omega_M$. We refer to this experimental condition as the slow-sampling scenario.

Our goal is to obtain necessary and sufficient conditions for the identifiability of $G_0(p)$, and to derive consistent nonparametric and parametric estimators for this sampling regime.

\begin{remark}
A similar analysis to the one presented in this letter can be performed if a zero-order hold device is present, where a discrete-time formulation, not covered here, would be preferred.
\end{remark}

\section{A least-squares estimate for the frequency response function}
\label{sec:leastsquares}
A nonparametric model for the system $G_0(p)$ can be obtained via the estimation of its frequency response function \cite{pintelon2012system}. Since the input in \eqref{input} only excites the frequencies $\omega=0,\pm\omega_1,\dots, \pm\omega_M$, we focus on estimating the frequency points contained in the vector
\begin{align}
    \mathbf{G}_0^\textnormal{f}: = \big[G_0(0)&, G_0(-\mathrm{i}\omega_1), G_0(\mathrm{i}\omega_1), \notag \\
    \label{g0f}
    &\dots, G_0(-\mathrm{i}\omega_M), G_0(\mathrm{i}\omega_M)\big]^\top\in \mathbb{C}^{2M+1}.
\end{align}
We assume the output is measured in a stationary regime, i.e., after all distortions caused by transient effects have decayed to zero. Then, the noiseless output $x(kh)$ is given by
\begin{align}
    x(\hspace{-0.01cm}kh\hspace{-0.01cm})\hspace{-0.065cm} &=\hspace{-0.055cm} G_{\hspace{-0.02cm}0}\hspace{-0.02cm}(0)a_{0} \hspace{-0.065cm}+\hspace{-0.075cm} \sum_{\ell =1}^M \hspace{-0.06cm}a_\ell |G_{\hspace{-0.02cm}0}\hspace{-0.02cm}(\hspace{-0.01cm}\mathrm{i}\omega_\ell\hspace{-0.015cm})| \hspace{-0.02cm}\cos\hspace{-0.07cm}\big(\omega_\ell kh \hspace{-0.06cm}+\hspace{-0.06cm} \phi_{\ell}\hspace{-0.06cm}+\hspace{-0.085cm}\angle G_{\hspace{-0.02cm}0}\hspace{-0.02cm}(\mathrm{i}\omega_\ell)\hspace{-0.01cm}\big)  \notag \\
    \label{linear}
    &= \bm{\zeta}^\hop(kh) \mathbf{G}_0^\textnormal{f},
\end{align}
where
\begin{equation}
\label{zeta}
    \bm{\zeta}(kh) =  \begin{bmatrix}
        a_{0} \\
        \frac{a_1}{2} e^{\mathrm{i}(\omega_1 kh +\phi_{1})} \\
        \frac{a_1}{2} e^{-\mathrm{i}(\omega_1 kh +\phi_{1})} \\
        \vdots \\
        \frac{a_M}{2} e^{\mathrm{i}(\omega_M kh +\phi_{M})} \\
        \frac{a_M}{2} e^{-\mathrm{i}(\omega_M kh +\phi_{M})}
    \end{bmatrix}.
\end{equation}
Thus, the following least-squares estimator to compute the frequency response vector $\mathbf{G}_0^\textnormal{f}$ is considered:
\begin{equation}
\label{ls}
    \hat{\mathbf{G}}^{\textnormal{f}}\hspace{-0.04cm}=\hspace{-0.05cm} \left[\sum_{k=1}^N \bm{\zeta}(kh) \bm{\zeta}^\hop(kh) \right]^{\hspace{-0.04cm}-1}\hspace{-0.1cm}\left[\sum_{k=1}^N  \bm{\zeta}(kh) y(kh) \right].
\end{equation}

\subsection{Statistical properties of the least-squares estimator}
For the least-squares estimator in \eqref{ls} to be well-defined in the slow-sampling scenario, we require the input frequencies $\{\omega_\ell\}_{\ell=1}^M$ to not overlap after accounting for aliasing. This key condition is made explicit in Assumption \ref{assumption1}.
\begin{assumption}[Non-overlapping condition]
\label{assumption1}
The input frequencies $\{\omega_\ell\}_{\ell=1}^M$ satisfy
    \begin{equation}
    \label{frequencycondition}
           \begin{cases}
               \hspace{-0.02cm}\omega_\ell\hspace{-0.04cm}\pm\hspace{-0.05cm} \omega_\tau \hspace{-0.06cm}\neq \hspace{-0.06cm} \frac{2 n \pi}{h} &\hspace{-0.22cm}\textnormal{for all } \ell,\tau= 1,\dots, M; \ell\neq \tau; n\in \mathbb{Z}, \\
               \hspace{-0.02cm}\omega_\ell\neq  \frac{n \pi}{h} &\hspace{-0.22cm}\textnormal{for all } \ell = 1,\dots, M; n\in \mathbb{Z}.
           \end{cases}
    \end{equation}
\end{assumption}
Note that for each pair of frequencies below the Nyquist frequency, the inequalities in \eqref{frequencycondition} are satisfied.

The well-posedness of the estimator defined in \eqref{ls} and its unbiasedness when the sampling period does not satisfy the Nyquist-Shannon criterion are investigated in Theorem \ref{thm1}. These results constitute Contribution \ref{contribution1} of this letter.
\begin{theorem}
\label{thm1}
    Consider the sampled input and output signals $\{u(kh),y(kh)\}_{k=1}^N$, where $u(t)$ is given by \eqref{input}, $y(kh)$ is measured in a stationary regime, $h\geq \pi/\omega_M$, and $N>2M$. Then, \eqref{ls} is well-defined and is an unbiased estimator of the frequency response vector $\mathbf{G}_0^{\textnormal{f}}$ if and only if Assumption \ref{assumption1} holds. In such case, its covariance is given by
    \begin{equation}
    \label{covariance}
    \textnormal{Cov}\{\hat{\mathbf{G}}^{\textnormal{f}}\} = \sigma^2 \left[ \sum_{k=1}^N \bm{\zeta}(kh) \bm{\zeta}^\hop(kh) \right]^{-1}.
\end{equation}
\end{theorem}
\hspace{0.7cm}\textit{Proof:} The least-squares estimator \eqref{ls} is well-defined if and only if its associated normal matrix 
\begin{equation}
\label{z}
    \mathbf{Z} :=\sum_{k=1}^N \bm{\zeta}(kh) \bm{\zeta}^\hop(kh)
\end{equation}
is nonsingular. Let $\mathbf{w}\in \mathbb{C}^{2M+1}$ be arbitrary. We have $\mathbf{w}^\hop\mathbf{Z}\mathbf{w}=0$ if and only if  $\mathbf{K}\mathbf{w}=\mathbf{0}$, where $\mathbf{K}=[\bm{\zeta}(h),\dots,\bm{\zeta}(Nh)]^\hop$. The matrix $\mathbf{K}$ can be decomposed as $\mathbf{K}=\mathbf{V} \textnormal{diag}(\bm{\zeta}^\hop(h))$, where $\mathbf{V}\in \mathbb{C}^{N\times (2M+1)}$ is a rectangular Vandermonde matrix given by
\begin{small}
    \begin{equation}
    \mathbf{V} \hspace{-0.1cm}=\hspace{-0.1cm} \begin{bmatrix}
        1  \hspace{-0.2cm}& 1 \hspace{-0.1cm}&\hspace{-0.1cm} 1 & \hspace{-0.2cm}\cdots \hspace{-0.2cm}& \hspace{-0.1cm}1\hspace{-0.1cm} & 1 \\
        1  \hspace{-0.2cm}& e^{-\mathrm{i}\omega_1 h} \hspace{-0.1cm}&\hspace{-0.1cm} e^{\mathrm{i}\omega_1 h} & \hspace{-0.2cm}\cdots \hspace{-0.2cm}& \hspace{-0.1cm} e^{-\mathrm{i}\omega_M h} \hspace{-0.1cm}& e^{\mathrm{i}\omega_M h} \\
        \vdots  \hspace{-0.2cm}& \vdots \hspace{-0.1cm}&\hspace{-0.1cm} \vdots & \hspace{-0.2cm}\cdots\hspace{-0.2cm} & \hspace{-0.1cm}\vdots\hspace{-0.1cm} & \vdots \\
        1  \hspace{-0.2cm}& e^{-\mathrm{i}\omega_1 (N\hspace{-0.03cm}-\hspace{-0.03cm}1)h} \hspace{-0.1cm}&\hspace{-0.1cm} e^{\mathrm{i}\omega_1 (N\hspace{-0.03cm}-\hspace{-0.03cm}1) h} & \hspace{-0.2cm}\cdots\hspace{-0.2cm} & e^{-\mathrm{i}\omega_M (N\hspace{-0.03cm}-\hspace{-0.03cm}1) h} \hspace{-0.1cm}& \hspace{-0.1cm}e^{\mathrm{i}\omega_M (N\hspace{-0.03cm}-\hspace{-0.03cm}1) h}
    \end{bmatrix}\hspace{-0.04cm}. \notag
\end{equation}
\end{small}
\hspace{-0.15cm}Vandermonde matrices such as the one above are known to have full column rank if $N\geq 2M+1$ and the complex numbers $1,e^{\pm \mathrm{i}\omega_1 h},\dots, e^{\pm \mathrm{i}\omega_M h}$ are distinct \cite[Sec. 0.9.11]{Horn2012}. This necessary and sufficient condition is equivalent to \eqref{frequencycondition}. Under the assumptions in \eqref{frequencycondition}, $\mathbf{K}$ has full column rank and thus $\mathbf{K}\mathbf{w}=\mathbf{0}$ implies $\mathbf{w}=\mathbf{0}$, which means that $\mathbf{Z}$ is positive definite and hence nonsingular.

For the unbiasedness result, we compute the expected value of the least-squares estimator in \eqref{ls} as
\begin{equation}
    \mathbb{E}\{\hat{\mathbf{G}}^{\textnormal{f}}\}=\mathbf{Z}^{-1}\left[\sum_{k=1}^N  \bm{\zeta}(kh)x(kh) \right] = \mathbf{G}_0^\textnormal{f}, \notag
\end{equation}
where the last equality is due to \eqref{linear}. Finally, by noting that
\begin{equation}
    \hat{\mathbf{G}}^{\textnormal{f}}-\mathbb{E}\{\hat{\mathbf{G}}^{\textnormal{f}}\} =\mathbf{Z}^{-1}\left[\sum_{k=1}^N  \bm{\zeta}(kh)v(kh) \right], \notag 
\end{equation}
we have, using the fact that $\{v(kh)\}$ is white noise,
\begin{align}
    \textnormal{Cov}\{\hat{\mathbf{G}}^{\textnormal{f}}\} &= \mathbf{Z}^{-1} \sum_{k=1}^N \sum_{\tau=1}^N \bm{\zeta}(kh)\mathbb{E}\big\{\hspace{-0.02cm} v(kh) v(\tau h)\hspace{-0.02cm}\big\}\bm{\zeta}^\hop(\tau h) \mathbf{Z}^{-1} \notag  \\
    &= \sigma^2 \mathbf{Z}^{-1} \sum_{k=1}^N \bm{\zeta}(kh)\bm{\zeta}^\hop(kh) \mathbf{Z}^{-1} \notag  \\
    &= \sigma^2 \mathbf{Z}^{-1}. \tag*{$\blacksquare$}
\end{align}
\begin{corollary}
\label{corollary31}
    Consider the same experimental conditions as in Theorem \ref{thm1}, and assume that Assumption \ref{assumption1} holds. Then, \eqref{ls} is a consistent estimator of $\mathbf{G}_0^{\textnormal{f}}$ in \eqref{g0f}. Furthermore, the frequency response estimates $\hat{G}(0), \hat{G}(\pm \mathrm{i}\omega_\ell),\ell = 1,\dots,M$ are asymptotically mutually uncorrelated, and the asymptotic covariance of $\hat{\mathbf{G}}^{\textnormal{f}}$ is given by
    \begin{equation}
    \label{asymptoticcovariance}
        \lim_{N\to\infty}  \hspace{-0.14cm}N \textnormal{Cov}\{\hspace{-0.01cm}\hat{\mathbf{G}}^{\textnormal{f}}\hspace{-0.01cm}\} \hspace{-0.08cm}=\hspace{-0.07cm} \sigma^2 \hspace{-0.01cm} \textnormal{diag}\hspace{-0.075cm}\left(\hspace{-0.06cm}\frac{1}{a_0^2}\hspace{-0.01cm}, \hspace{-0.025cm}\frac{4}{a_1^2}, \hspace{-0.025cm}\frac{4}{a_1^2},\hspace{-0.025cm} \dots\hspace{-0.01cm},\hspace{-0.03cm} \frac{4}{a_M^2}, \hspace{-0.025cm}\frac{4}{a_M^2}\hspace{-0.05cm}\right)\hspace{-0.06cm}.
    \end{equation}
\end{corollary}
\begin{proof}
An estimator is consistent if it is unbiased and its covariance decays to zero \cite[p. 54]{lehmann1998theory}. Unbiasedness follows from Theorem \ref{thm1}. We now compute its asymptotic covariance. From \eqref{covariance}, the asymptotic covariance is given by $\sigma^2\bar{\mathbf{Z}}^{-1}$, where $\bar{\mathbf{Z}} = \lim_{N\to \infty}\mathbf{Z}/N$, with $\mathbf{Z}$ being defined in \eqref{z}. Calculating the entries of $\bar{\mathbf{Z}}_{ij}$ of $\bar{\mathbf{Z}}$, we find $\bar{\mathbf{Z}}_{11}=a_0^2$ and $\bar{\mathbf{Z}}_{(2\ell)(2\ell)}=\bar{\mathbf{Z}}_{(2\ell+1)(2\ell+1)}=a_{\ell}^2/4$ for $\ell\in \{1,2,\dots, M\}$. Lastly, every non-diagonal element of $\bar{\mathbf{Z}}$ is of the form
	\begin{equation}
 \label{limit}
	\lim_{N\to \infty} \frac{C}{N}\sum_{k=1}^N e^{\mathrm{i}(\tilde{\omega}kh+\tilde{\psi})}
	\end{equation}
	for some real constants $C,\tilde{\psi}$. Due to the condition in \eqref{frequencycondition}, the constant $\tilde{\omega}$, which is the difference or sum of frequencies $\omega_\ell$ (including the zero frequency), is not a multiple of $2\pi/h$. The following chain of inequalities is satisfied for $N\geq 2M+1$:
	\begin{equation}
	0\hspace{-0.05cm}\leq \hspace{-0.06cm}\left| \hspace{-0.02cm}\frac{C}{N} \hspace{-0.07cm}\sum_{k=1}^N \hspace{-0.06cm}e^{\mathrm{i}(\tilde{\omega}kh\hspace{-0.01cm}+\hspace{-0.01cm}\tilde{\psi})} \hspace{-0.02cm}\right| \hspace{-0.08cm}=\hspace{-0.08cm} \left|\hspace{-0.02cm}\frac{C\hspace{-0.04cm}\sin(\frac{1}{2}\hspace{-0.01cm}[N\hspace{-0.09cm}+\hspace{-0.07cm}1]\tilde{\omega} h)}{N\sin(\frac{1}{2}\tilde{\omega} h)}\hspace{-0.02cm}\right|\hspace{-0.055cm} \leq\hspace{-0.055cm} \left|\hspace{-0.02cm}\frac{C}{N\hspace{-0.045cm}\sin(\frac{1}{2}\tilde{\omega} h)}\hspace{-0.02cm}\right|\hspace{-0.03cm}. \notag
	\end{equation}
	Since $\tilde{\omega}\neq 2n\pi/h$, $n\in \mathbb{Z}$, the denominator of the upper bound is nonzero. The squeeze theorem \cite[Theorem 3.3.6]{sohrab2003basic} then implies that the limit in \eqref{limit} is equal to zero. This means that all the off-diagonal elements of $\bar{\mathbf{Z}}$ are zero, leading to \eqref{asymptoticcovariance}. Finally, since the covariance of \eqref{ls} decays to zero, it is consistent, concluding the proof.
\end{proof}

In summary, unbiased estimates of the frequency response function beyond the Nyquist frequency $\pi/h$ can be obtained if the input frequencies do not overlap after aliasing, as indicated by the condition in \eqref{frequencycondition}. Frequencies equal to integer multiples of the Nyquist frequency, i.e., $\omega_\ell = n\pi/h$, alias to the zero frequency if $n$ is even, or a single frequency line if $n$ is odd. In the latter case, after minor adjustments to the proof of Theorem \ref{thm1}, it can be shown that only the real part of $G_0(\mathrm{i}n\pi/h)$ can be estimated unbiasedly. Note that only one experiment is needed for the least-squares estimator in \eqref{ls} to be computed, and no assumptions on the smoothness of the frequency response of the system are required.

\subsection{Frequency-domain interpretation}
The least-squares estimator in \eqref{ls} has a frequency-domain interpretation. Via Parseval's theorem \cite{oppenheim1997signals}, \eqref{ls} is equivalent~to
\begin{equation}
\label{lsfreq}
    \hat{\mathbf{G}}^{\textnormal{f}}\hspace{-0.07cm}=\hspace{-0.08cm} \left[\sum_{n=1}^N \bm{\Psi}[e^{\mathrm{i}\frac{2\pi n}{N}}] \bm{\Psi}^\hop[e^{\mathrm{i}\frac{2\pi n}{N}}] \right]^{\hspace{-0.04cm}-\hspace{-0.02cm}1}\hspace{-0.1cm}\left[\sum_{n=1}^N  \bm{\Psi}[e^{\mathrm{i}\frac{2\pi n}{N}}] Y[e^{-\mathrm{i}\frac{2\pi n}{N}}] \right]\hspace{-0.04cm},
\end{equation}
where the discrete-time Fourier transforms (DTFTs) of $\bm{\zeta}(kh)$ and $y(kh)$ are respectively given by
\begin{equation}
    \label{dtfty}
    \bm{\Psi}[e^{\mathrm{i}\omega h}] = \sum_{k=1}^N \bm{\zeta}(kh)e^{-\mathrm{i}hk\omega}, \hspace{0.2cm} Y[e^{\mathrm{i}\omega h}] = \sum_{k=1}^N y(kh)e^{-\mathrm{i}hk\omega}.
\end{equation}
To obtain an explicit formula for $\hat{\mathbf{G}}^{\textnormal{f}}$, we introduce the following assumption.
\begin{assumption}[No spectral leakage]
\label{assumption2}
There is no spectral leakage when computing $\bm{\Psi}[e^{i\omega h}]$, i.e., $Nh$ is a multiple of the least common multiple of $\{2\pi/\omega_\ell\}_{\ell=1}^M$.     
\end{assumption}
\begin{theorem}
\label{thm32}
    Consider the sampled input and output signals $\{u(kh),y(kh)\}_{k=1}^N$, where $u(t)$ is given by \eqref{input}, $y(kh)$ is measured in a stationary regime, $h\geq \pi/\omega_M$, and $N>2M$. Assume that Assumptions \ref{assumption1} and \ref{assumption2} hold. Then, the least-squares estimator \eqref{ls} can be computed as
    \begin{align}
    \hat{\mathbf{G}}^\textnormal{f} = \Bigg[\frac{Y[e^{\mathrm{i}0}]}{U[e^{\mathrm{i}0}]}&, \frac{Y[e^{-\mathrm{i}\omega_1 h}]}{U[e^{-\mathrm{i}\omega_1 h}]}, \frac{Y[e^{\mathrm{i}\omega_1 h}]}{U[e^{\mathrm{i}\omega_1 h}]}, \notag \\
    \label{lsfreq2}
    &\dots, \frac{Y[e^{-\mathrm{i}\omega_M h}]}{U[e^{-\mathrm{i}\omega_M h}]}, \frac{Y[e^{\mathrm{i}\omega_M h}]}{U[e^{\mathrm{i}\omega_M h}]}\Bigg]^\top,
\end{align}
where the DTFT of $y(kh)$ is given by \eqref{dtfty}, and
\begin{equation}
    U[e^{\mathrm{i}0}] =Na_0, U[e^{\pm\mathrm{i}\omega_\ell h}] =N\frac{a_\ell}{2}e^{\pm\mathrm{i}\phi_\ell},\ell=1,\dots, M. \notag
\end{equation}
Furthermore, the frequency response estimates $\hat{G}(0), \hat{G}(\pm\mathrm{i}\omega_\ell),\ell=1,\dots, M$ are mutually uncorrelated for any $N\geq 2M+1$.
\end{theorem}
\begin{proof}
The discrete Fourier transform of $\bm{\zeta}(kh)$ is a $N$-periodic function described in one period by
\begin{equation}
\label{modexpression}
    \bm{\Psi}[e^{\mathrm{i}\frac{2\pi n}{N}}]\hspace{-0.05cm} =\hspace{-0.05cm} N \hspace{-0.04cm}\begin{bmatrix}
        a_{0}\delta\hspace{-0.05cm}\left(n\textnormal{ mod } N \right) \\
        \frac{a_1}{2} e^{\mathrm{i}\phi_{1}}\delta\hspace{-0.05cm}\left((n-\frac{Nh\omega_1}{2\pi})\textnormal{ mod } N \right) \\
        \frac{a_1}{2} e^{-\mathrm{i}\phi_{1}}\delta\hspace{-0.05cm}\left((n+\frac{Nh\omega_1}{2\pi})\textnormal{ mod } N \right) \\
        \vdots \\
        \frac{a_M}{2} e^{\mathrm{i}\phi_{M}}\delta\hspace{-0.05cm}\left((n-\frac{Nh\omega_M}{2\pi})\textnormal{ mod } N \right) \\
        \frac{a_M}{2} e^{-\mathrm{i}\phi_{M}}\delta\hspace{-0.05cm}\left((n+\frac{Nh\omega_M}{2\pi})\textnormal{ mod } N \right)
    \end{bmatrix},
\end{equation}
where we have used Assumption \ref{assumption2}. Since \eqref{frequencycondition} also holds, the matrix being inverted in \eqref{lsfreq} admits the expression
\begin{equation}
\label{combining1}
    \sum_{n=1}^N \hspace{-0.065cm}\bm{\Psi}[\hspace{-0.01cm}e^{\mathrm{i}\frac{2\pi n}{N}}\hspace{-0.01cm}] \bm{\Psi}^\hop[e^{\mathrm{i}\hspace{-0.02cm}\frac{2\pi n}{N}}\hspace{-0.01cm}]\hspace{-0.075cm} = \hspace{-0.085cm}N^2\textnormal{diag}\hspace{-0.07cm}\left(\hspace{-0.055cm}a_0^2, \hspace{-0.025cm}\frac{a_1^2}{4}, \hspace{-0.025cm}\frac{a_1^2}{4},\hspace{-0.025cm} \dots,\hspace{-0.025cm} \frac{a_M^2}{4}, \hspace{-0.025cm}\frac{a_M^2}{4}\hspace{-0.05cm}\right)\hspace{-0.06cm}. 
\end{equation}
Via Parseval's theorem applied to \eqref{combining1}, we find that $\mathbf{Z}$ in \eqref{z} is also a diagonal matrix, and thus the correlation between frequency response estimates is zero.

Similarly, using \eqref{modexpression} and the $N$-periodicity of $Y[e^{\mathrm{i}\frac{2\pi n}{N}}]$,
\begin{equation}
\label{combining2}
    \sum_{n=1}^N  \bm{\Psi}[e^{\mathrm{i}\frac{2\pi n}{N}}] Y[e^{-\mathrm{i}\frac{2\pi n}{N}}] = N \begin{bmatrix}
        a_{0} Y[e^{\mathrm{i}0}] \\
        \frac{a_1}{2} e^{\mathrm{i}\phi_{1}}Y[e^{-\mathrm{i}\omega_1 h}] \\
        \frac{a_1}{2} e^{-\mathrm{i}\phi_{1}}Y[e^{\mathrm{i}\omega_1 h}] \\
        \vdots \\
        \frac{a_M}{2} e^{\mathrm{i}\phi_{M}}Y[e^{-\mathrm{i}\omega_M h}] \\
        \frac{a_M}{2} e^{-\mathrm{i}\phi_{M}}Y[e^{\mathrm{i}\omega_M h}] 
    \end{bmatrix}.
\end{equation}
The computation of \eqref{lsfreq} via \eqref{combining1} and \eqref{combining2} leads to \eqref{lsfreq2}, which is what we aimed to prove.
\end{proof}

Theorem \ref{thm32} states that an unbiased estimate of the frequency response at the frequencies above the Nyquist frequency can be computed directly from the empirical transfer function estimate (ETFE) \cite{ljung1985estimation}, evaluated at the aliased frequencies. This insight provides a theoretical basis for interpreting the ETFE beyond the Nyquist frequency when the input suffers from aliasing.
\begin{remark}
The proof of Theorem \ref{thm32} shows that $\mathbf{Z}$ in \eqref{z} is a diagonal matrix when Assumption \ref{assumption2} holds. Thus, for a fixed sample size, the variance of the frequency response function estimates does not increase with increasing sampling periods, as long as no spectral leakage is introduced.
\end{remark}
The frequency response function estimator in \eqref{ls} or \eqref{lsfreq2} can be used to compute parametric estimators that are accurate above the Nyquist frequency, as seen next.

\section{Nonparametric to parametric estimators}
\label{sec:parametric}
Once the frequency response function estimate \eqref{ls} has been computed, we propose a parametric model $G(p,\bm{\theta})$ for the system $G_0(p)$, where $\bm{\theta}\in\mathbb{R}^{n_\theta}$ contains the coefficients associated with, e.g., numerator and denominator polynomials of an unfactored rational transfer function, or coefficients associated with a modal parametrization of the system \cite{gonzalez2024identification}. Throughout this section, we assume that there exists a unique vector $\bm{\theta}_0\in \mathbb{R}^{n_\theta}$ such that $G(p,\bm{\theta}_0)=G_0(p)$.

For the sequel, we define the vector of frequency responses of the parametric model as
\begin{align}
    \mathbf{G}^\textnormal{f}(\bm{\theta})&: = \big[G(0,\bm{\theta}), G(-\mathrm{i}\omega_1,\bm{\theta}), G(\mathrm{i}\omega_1,\bm{\theta}), \notag \\
\label{Gftheta}
    &\qquad \dots, G(-\mathrm{i}\omega_M,\bm{\theta}), G(\mathrm{i}\omega_M,\bm{\theta})\big]^\top.
\end{align}
A parametric model can be obtained from the least-squares estimate \eqref{ls} by curve fitting in the frequency domain. This method, when weighted correctly, corresponds exactly to the time-domain prediction error method with a multisine predictor that suffers from aliasing. This result, contained in Theorem \ref{thm2}, constitutes Contribution \ref{contribution2} of this letter.
\begin{theorem}
\label{thm2}
   Consider the sampled input and output signals $\{u(kh),y(kh)\}_{k=1}^N$, where $u(t)$ is given by \eqref{input}, $y(kh)$ is measured in a stationary regime, $h\geq \pi/\omega_M$, $N>2M$, and assume that Assumption \ref{assumption1} holds. The following two cost functions share the same global minima:
    \begin{align}
        \label{vf}
        V_\textnormal{f}(\bm{\theta}) &=\hspace{-0.04cm} \big(\mathbf{G}^\textnormal{f}(\bm{\theta})\hspace{-0.04cm}-\hspace{-0.04cm}\hat{\mathbf{G}}^\textnormal{f}\big)^\hop \big[\textnormal{Cov}\{\hat{\mathbf{G}}^\textnormal{f}\}\big]^{-1}\big(\mathbf{G}^\textnormal{f}(\bm{\theta})\hspace{-0.04cm}-\hspace{-0.04cm}\hat{\mathbf{G}}^\textnormal{f}\big), \\
        \label{vt}
        V_\textnormal{t}(\bm{\theta}) &= \sum_{k=1}^N \big(y(kh)-\hat{y}(kh,\bm{\theta}) \big)^2,
    \end{align}
    where $\hat{\mathbf{G}}^{\textnormal{f}}$ is from \eqref{ls}, and the predictor $\hat{y}(kh,\bm{\theta})$ is given by
    \begin{align}
        \hat{y}(kh,&\bm{\theta}) = G(0,\bm{\theta})a_{0} \notag \\
        \label{predictory}
        &\hspace{-0.7cm}+ \sum_{\ell =1}^M a_\ell |G(\mathrm{i}\omega_\ell,\bm{\theta})| \cos\big(\omega_\ell kh + \phi_{\ell}+\angle G(\mathrm{i}\omega_\ell,\bm{\theta})\big).
    \end{align}    
\end{theorem}
\begin{proof}
    When analyzed as a function of $\bm{\theta}$, except for an additive constant, $V_{\textnormal{f}}(\bm{\theta})$ is proportional to 
    \begin{equation}
    \label{combine0}
        V_{\textnormal{f}}(\bm{\theta})\propto [\mathbf{G}^{\textnormal{f}}(\bm{\theta})]^\hop \mathbf{Z} \mathbf{G}^{\textnormal{f}}(\bm{\theta}) - 2\textnormal{Re}\big\{ (\hat{\mathbf{G}}^{\textnormal{f}})^\hop \mathbf{Z} \mathbf{G}^{\textnormal{f}}(\bm{\theta}) \big\},
    \end{equation}
    where we used the fact that $\textnormal{Cov}\{\hat{\mathbf{G}}^\textnormal{f}\} = \sigma^2 \mathbf{Z}^{-1}$, with $\mathbf{Z}$ defined as in \eqref{z}. First, by leveraging the result in \eqref{linear},  
    \begin{equation}
        [\mathbf{G}^{\textnormal{f}}(\bm{\theta})]^\hop \mathbf{Z} \mathbf{G}^{\textnormal{f}}(\bm{\theta}) = \sum_{k=1}^N |\bm{\zeta}^\hop (kh)\mathbf{G}^{\textnormal{f}}(\bm{\theta})|^2 = \sum_{k=1}^N \hat{y}^2(kh,\bm{\theta}).   \notag
    \end{equation}
On the other hand, the identity $(\hat{\mathbf{G}}^{\textnormal{f}})^\hop\mathbf{Z} = \sum_{k=1}^N  y(kh) \bm{\zeta}^\hop(kh)$ leads to
\begin{equation}
    \textnormal{Re}\big\{ (\hat{\mathbf{G}}^{\textnormal{f}})^\hop \mathbf{Z} \mathbf{G}^{\textnormal{f}}(\bm{\theta}) \big\} = \sum_{k=1}^N  y(kh)\hat{y}(kh,\bm{\theta}). \notag
\end{equation}
Replacing $[\mathbf{G}^{\textnormal{f}}(\bm{\theta})]^\hop \mathbf{Z} \mathbf{G}^{\textnormal{f}}(\bm{\theta})$ and $\textnormal{Re}\{ (\hat{\mathbf{G}}^{\textnormal{f}})^\hop \mathbf{Z} \mathbf{G}^{\textnormal{f}}(\bm{\theta}) \}$ in \eqref{combine0} and completing the square leads to $V_\textnormal{f}(\bm{\theta})\propto V_{\textnormal{t}}(\bm{\theta})$. Thus, these cost functions share the same global minima.
\end{proof}

Theorem \ref{thm2} indicates that, provided the input frequency conditions in \eqref{frequencycondition} are satisfied and $N\geq 2M+1$, there is an equivalence between frequency domain and time domain parametric estimation for the slow-sampling scenario. The expressions also coincide with the standard prediction error method and maximum likelihood estimators for the Gaussian noise case when the sampling rate is higher than the Nyquist frequency \cite{ljung1998system}. The parametric model can be obtained via numerical minimization of \eqref{vf} or \eqref{vt}. For example, the minimization of \eqref{vt} can be achieved using Gauss-Newton (GN) iterations \cite[Sec. 10.2]{ljung1998system}:
\begin{equation}
    \bm{\theta}_{j+1} \hspace{-0.05cm}=\hspace{-0.05cm} \bm{\theta}_j \hspace{-0.05cm}+\hspace{-0.05cm} \left[\sum_{k=1}^N \hat{\bm{\varphi}}_j(kh)\hat{\bm{\varphi}}_j^\top\hspace{-0.03cm}(kh)\right]^{\hspace{-0.03cm}-1}\hspace{-0.05cm}\left[\sum_{k=1}^N \hat{\bm{\varphi}}_j(kh)\varepsilon_j(kh)\hspace{-0.02cm}\right], \notag 
\end{equation}
where $\hat{\bm{\varphi}}_j(kh)=(\partial \hat{y}(kh,\bm{\theta})/\partial \bm{\theta})|_{\bm{\theta}=\bm{\theta}_j}$ and $\varepsilon_j(kh) = y(kh)-\hat{y}(kh,\bm{\theta}_j)$, with $\hat{y}(kh,\bm{\theta}_j)$ being defined in \eqref{predictory}. As in the standard sampling case $h < \pi/\omega_M$, caution must be taken when initializing the iterative procedure, as convergence to a global minimum is not guaranteed \cite{ljung1998system}.

\begin{remark}
    Under Assumption \ref{assumption2} on the absence of spectral leakage, applying the result in \eqref{combining1} in $V_{\textnormal{f}}(\bm{\theta})$ gives
    \begin{equation}
        V_{\textnormal{f}}(\bm{\theta}) \hspace{-0.05cm} \propto \hspace{-0.03cm} a_0^2 \big|\hat{G}(0)\hspace{-0.04cm}-\hspace{-0.04cm}G(0,\bm{\theta})\big|^2 \hspace{-0.03cm}+\hspace{-0.03cm} \sum_{\ell=1}^M \hspace{-0.03cm}\frac{a_\ell^2}{2} \big|\hat{G}(\mathrm{i}\omega_\ell)\hspace{-0.04cm}-\hspace{-0.04cm}G(\mathrm{i}\omega_\ell,\bm{\theta})\big|^2, \notag 
    \end{equation}
    i.e., the least-squares cost function of the frequency response estimates, weighted by the magnitude of the input spectrum.
\end{remark}

In the slow-sampling scenario, the prediction error method proposed in Theorem \ref{thm2} provides consistent estimators of the parametric model $G(p,\bm{\theta})$ under mild conditions. In other words, the parametric estimator that minimizes \eqref{vf} and \eqref{vt} converges to $\bm{\theta}_0$ with probability 1. Interestingly, this property holds even if the aliasing constraint in Assumption \ref{assumption1} is not satisfied, i.e., if there are input frequencies that overlap after aliasing. Theorem \ref{thmconsistency1} formalizes this result, which is Contribution \ref{contribution3} of this letter.

\begin{theorem}
    \label{thmconsistency1}
   Consider the sampled input and output signals $\{u(kh),y(kh)\}_{k=1}^N$, where $u(t)$ is given by \eqref{input}, $y(kh)$ is measured in a stationary regime, and $h\geq \pi/\omega_M$.  Assume that the model satisfies the condition
    \begin{equation}
    \label{identifiability}
\mathbf{G}^\textnormal{f}(\bm{\theta})-\mathbf{G}_0^\textnormal{f} \in \textnormal{Ker}(\bm{\Phi}_{\bm{\zeta}}) \implies 
        \bm{\theta}=\bm{\theta}_0,
    \end{equation}
    where $\bm{\Phi}_{\bm{\zeta}} :=\overline{\mathbb{E}}\{\bm{\zeta}(kh)\bm{\zeta}^\hop(kh)\}$. Then, as $N$ tends to infinity, the parametric estimator that minimizes \eqref{vt} among all asymptotically stable models converges to $\bm{\theta}_0$ with probability 1.
\end{theorem}
\begin{proof}
By Lemma 8.2 of \cite{ljung1998system}, the criterion $V_\textnormal{t}(\bm{\theta})/N$ converges uniformly and with probability 1 as $N\to\infty$ to
\begin{equation}
    \overline{V}(\bm{\theta}) = \overline{\mathbb{E}}\big\{\big(y(kh)-\hat{y}(kh,\bm{\theta})\big)^2 \big\}. \notag 
\end{equation}
Due to \eqref{linear} and $\hat{y}(kh,\bm{\theta})=\bm{\zeta}^\hop(kh) \hat{\mathbf{G}}^\textnormal{f}(\bm{\theta})$, we can compute
\begin{align}
    \hspace{-0.4cm}\overline{V}(\bm{\theta}) &= \overline{\mathbb{E}}\big\{\big(x(kh)-\hat{y}(kh,\bm{\theta})\big)^2 \big\} + \sigma^2 \notag \\
    &= \lim_{N\to \infty}\frac{1}{N}\sum_{k=1}^N \left| \bm{\zeta}^\hop(kh)(\mathbf{G}^{\textnormal{f}}(\bm{\theta})-\mathbf{G}_0^{\textnormal{f}}) \right|^2 + \sigma^2 \notag \\
    &=(\mathbf{G}^{\textnormal{f}}(\bm{\theta})-\mathbf{G}_0^{\textnormal{f}})^\hop \bm{\Phi}_{\bm{\zeta}}(\mathbf{G}^{\textnormal{f}}(\bm{\theta})-\mathbf{G}_0^{\textnormal{f}}) + \sigma^2. \notag
\end{align}
This cost function is bounded below by $\sigma^2$, with equality if and only if $\bm{\Phi}_{\bm{\zeta}}(\mathbf{G}^{\textnormal{f}}(\bm{\theta})-\mathbf{G}_0^{\textnormal{f}}) = \mathbf{0}$. By the identifiability condition in \eqref{identifiability}, the minimizer of $\overline{V}(\bm{\theta})$ is unique and is equal to $\bm{\theta}_0$, concluding the proof.
\end{proof}

Following similar steps from the proof of Corollary \ref{corollary31}, the matrix $\bm{\Phi}_{\bm{\zeta}}$ has a rank equal to the number of distinct input frequency lines (after accounting for aliasing) within the fundamental frequency band $[-\pi/h, \pi/h)$. If \eqref{frequencycondition} is satisfied, $\bm{\Phi}_{\bm{\zeta}}$ is nonsingular, and \eqref{identifiability} reduces to the standard identifiability condition $\mathbf{G}^\textnormal{f}(\bm{\theta}) = \mathbf{G}_0^\textnormal{f} \implies \bm{\theta} = \bm{\theta}_0$. Therefore, when no input frequencies overlap after aliasing, $n_\theta \leq 2M + 1$ ensures identifiability for standard model parametrizations and consistency of the prediction error method estimator. If input frequency lines overlap, consistency is still guaranteed as long as the dimension of $\bm{\theta}$ does not exceed the number of unique non-overlapping input frequency lines.

\section{Simulation studies}
\label{sec:simulations}
In this section we verify the main theoretical findings via Monte Carlo simulations. The tests are conducted on a benchmark system known as the Rao-Garnier system \cite{rao2002numerical}:
\begin{equation}
\label{raogarnier}
    G_0(p) = \frac{-6400p+1600}{p^4+5p^3+408p^2+416p+1600},
\end{equation}
with $\bm{\theta}_0 = [1600, 416,408,5,1600,-6400]^\top$ as true parameter vector. The parametric model we consider is given by
\begin{equation}
    G(p,\bm{\theta}) = \frac{b_1p+b_0}{p^4+a_3p^3+a_2 p^2+a_1 p+a_0}, \notag
\end{equation}
where $\bm{\theta} = [a_0,a_1,a_2,a_3,b_0,b_1]^\top$. The system in \eqref{raogarnier} is excited with an input as in \eqref{input}, and the noise variance corresponds to an output signal-to-noise ratio of $10$~[dB].

\subsection{Nonparametric system identification}
\label{subsec:nonparametric}
To assess the statistical performance of the nonparametric estimator in \eqref{ls}, we conduct 2000 Monte Carlo runs with varying noise realizations. Each simulation generates $N=2000$ output data samples using a fixed input with unit amplitudes, random phases, and $M=13$ nonzero frequencies between $0.1$ and $30$~[rad/s]. The sampling period is set to $h=0.5$~[s], which is 100 times larger than the typical sampling period for this system \cite{rao2002numerical}. Assumption \ref{assumption1} is satisfied under these experimental conditions.

\begin{figure}
	\centering{
		\includegraphics[width=0.48\textwidth]{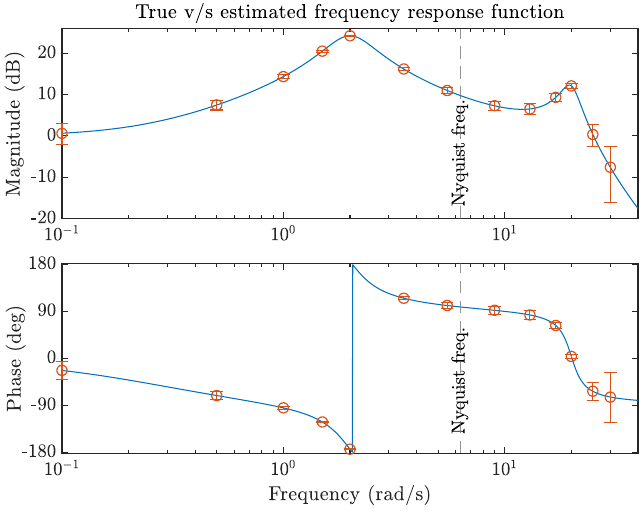}
  \vspace{-0.4cm}
		\caption{Bode plot of the system (blue), and the mean value of the magnitude and phase of the estimated frequency response function via least-squares, with its $95\%$ confidence interval (red). The least-squares approach provides an unbiased estimate of the system over the Nyquist frequency.}
		\label{fig1}}
  \vspace{-0.3cm}
\end{figure}

Figure~\ref{fig1} shows the Bode plot of the true system and the mean value of the frequency response estimate at each input frequency. As a metric of dispersion of the estimates, we have also plotted the 95\% confidence intervals associated with the magnitude and phase of the frequency response estimate. The estimator exhibits no noticeable bias at any frequency, including those above the Nyquist frequency. 

\subsection{Parametric system identification}
\label{subsec:parametric}
In this study we verify the consistency of the estimator derived from minimizing $V_{\textnormal{t}}(\bm{\theta})$ in \eqref{vt}. Specifically, we examine the sample mean and mean-square error (MSE) of the estimated parameters across 60 logarithmically spaced sample sizes ranging from $N=2\cdot10^3$ to $2\cdot 10^5$. For each $N$, 1000 Monte Carlo runs are performed using a fixed random-phase multisine input with unit amplitudes and nonzero frequencies $\pi/3, \pi,  7\pi/2$, and $5\pi$~[rad/s]. The frequency $5\pi$~[rad/s] overlaps with $\pi$ [rad/s], violating Assumption \ref{assumption1}. The maximum number of GN iterations is set to $100$, and the initial parameter vector is randomly generated, deviating from the true parameter vector by at most $10\%$ in each entry.

\begin{figure}
	\centering{
		\includegraphics[width=0.485\textwidth]{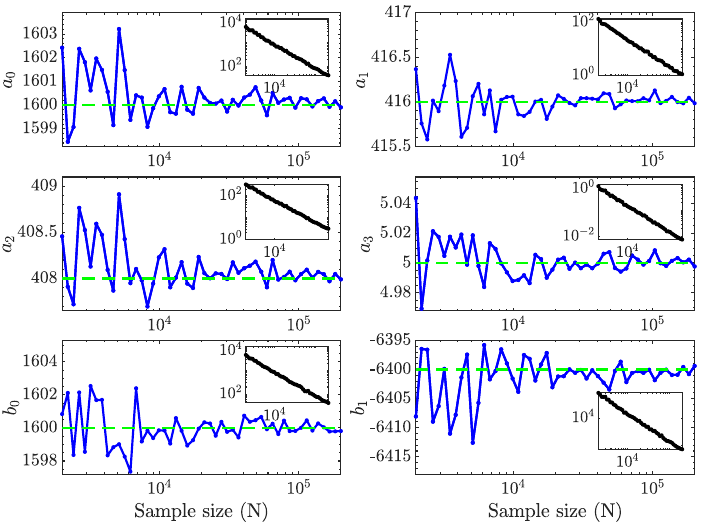}
    \vspace{-0.4cm}
		\caption{Empirical means of the parameter estimates (blue), with corresponding MSEs (log-log subplots, black lines). Parameters $a_i$ and $b_i$ correspond to the denominator and numerator coefficients of $G(p, \bm{\theta})$, respectively. All estimates converge to their true values (green dashed lines), and MSEs decay as $1/N$, indicating the consistency of the estimator minimizing \eqref{vt}.}
		\label{fig2}}
    \vspace{-0.3cm}
\end{figure}

The sample means for increasing sample sizes are reported in Figure~\ref{fig2}, where each plot contains a subplot with the MSE of each estimated parameter. For increasing values of $N$, all the parameter estimates converge to their true values, and the sample MSEs decay to zero as $1/N$. The estimator is consistent despite the input suffering from aliasing and frequency line overlapping, since the identifiability condition in \eqref{identifiability} applied to this example indicates that a maximum number of $7$ parameters can be estimated consistently. These results align with the findings in Theorem~\ref{thmconsistency1}.

\section{Conclusions}
\label{sec:conclusions}
The results in this letter provide statistical guarantees for nonparametric and parametric identification methods for slow-sampled continuous-time systems. Unbiased frequency response estimates can be obtained if the input frequencies avoid overlap after aliasing. For a fixed sample size and no spectral leakage, the covariance of these estimates does not deteriorate for increasing sampling periods. For parametric modelling, the prediction error method is consistent if the number of frequency lines below the Nyquist sampling rate is not less than the number of model parameters. Our methods and proofs extend to standard sampling rates ($h < \pi/\omega_M$), where aliasing is absent, aligning with standard identifiability conditions for the prediction error method \cite{ljung1998system}. Future work includes extending these results to MIMO system identification, where multiple experiments must be considered to identify the frequency response function at each input frequency \cite{pintelon2012system}.

\bibliography{references}
\end{document}